\pdfoutput=1
\documentclass[english]{lni}
\usepackage{float}
\usepackage{xspace}
\usepackage{url}
\RequirePackage{type1cm}
\RequirePackage{soul}
\setstcolor{blue}

\newcommand{\bscom}[3][]{%
   \noindent
   \st{#2}{\color{blue}\fontsize{8}{8}\selectfont\,#3}%
\ifx#1\empty\else{\color{red}\fontsize{8}{8}\selectfont\,[#1]}\fi
   }

\usepackage{graphicx}
\DeclareGraphicsExtensions{.png,.pdf,.ai,.jpg}
\setkeys{Gin}{pagebox=artbox}
\graphicspath{{./informatik20-figures/}}
\makeatletter
\def\input@path{{./informatik20-paper-invited/}}
\makeatother

\newcommand{\alphaweb}{$\alphaup$-web\xspace}
\newcommand{\betaweb}{$\betaup$-web\xspace}
\newcommand{\gammaweb}{$\gammaup$-web\xspace}
\newcommand{\deltaweb}{$\deltaup$-web\xspace}
\newcommand{\epsilonweb}{$\upvarepsilon$-web\xspace}

\newcommand{\args}{\href{https://args.me}{args.me}\xspace}
\newcommand{\chatnoir}{\href{https://www.chatnoir.eu}{ChatNoir}\xspace}
\newcommand{\netspeak}{\href{https://netspeak.org}{Netspeak}\xspace}
\newcommand{\picapica}{\href{https://picapica.org}{Picapica}\xspace}
\newcommand{\tira}{\href{https://www.tira.io}{Tira}\xspace}

\begin{document}

\title[Web Archive Analytics]{Web Archive Analytics\texorpdfstring{$^*$}{}}

\author[M. V\"olske et al.]{%
Michael~V\"olske\texorpdfstring{$^1$}{}
\and
Janek Bevendorff\texorpdfstring{$^1$}{}
\and
Johannes Kiesel\texorpdfstring{$^1$}{}
\and
Benno Stein\texorpdfstring{$^1$}{}
\and\qquad\qquad
Maik~Fr\"obe\texorpdfstring{$^2$}{}
\and
Matthias Hagen\texorpdfstring{$^2$}{}
\and
Martin~Potthast\texorpdfstring{$^3$}{}
}

\startpage{1} 
\editor{Names of the editors}
\booktitle{Informatik 2020} 
\year{2020}
\maketitle

\begin{abstract}
Web archive analytics is the exploitation of publicly accessible web pages and their evolution for research purposes---to the extent organizationally possible for researchers. In order to better understand the complexity of this task, the first part of this paper puts the entirety of the world's captured, created, and replicated data (the ``Global Datasphere'') in relation to other important data sets such as the public internet and its web pages, or what is preserved thereof by the Internet Archive. \\
Recently, the Webis research group, a network of university chairs to which the authors belong, concluded an agreement with the Internet Archive to download a substantial part of its web archive for research purposes. The second part of the paper in hand describes our infrastructure for processing this data treasure: We will eventually host around 8\,PB of web archive data from the Internet Archive and Common Crawl, with the goal of supplementing existing large scale web corpora and forming a non-biased subset of the 30\,PB web archive at the Internet Archive.
\end{abstract}


\makeatletter
\def\bsfootnote{\xdef\@thefnmark{$^{*}$\kern-0.2ex}\@footnotetext}
\makeatother

\makeatletter
\bsfootnote{An abstract of this paper has been published in the proceedings of the Open Search Symposium 2020.}
\footnotetext[1]{Bauhaus-Universit\"at Weimar, Germany. \email{<first>.<last>@uni-weimar.de}}
\footnotetext[2]{Martin-Luther-Universit\"at Halle-Wittenberg, Germany. \email{<first>.<last>@informatik.uni-halle.de}}
\footnotetext[3]{Leipzig University, Germany. \email{martin.potthast@uni-leipzig.de}}
\makeatother

\setcounter{footnote}{3}

\section{The Global Datasphere}

\begin{figure}[t]
\centering
\includegraphics[width=\textwidth]{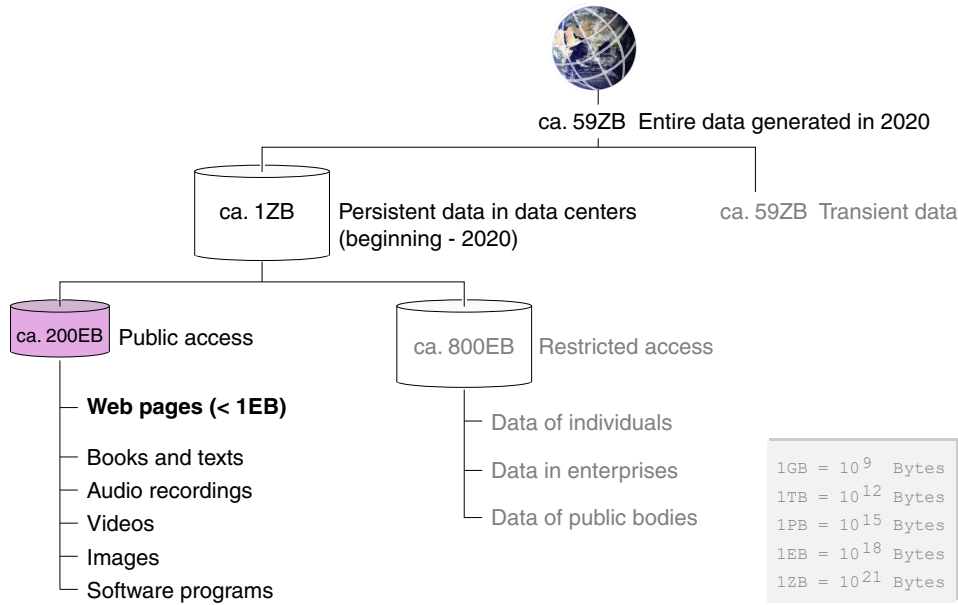}
\smallskip
\caption{Overview of the Global DataSphere in 2020. The shown numbers are estimates based on analyses and market forecasts by Cisco Systems, the International Data Corporation, and Statista Inc., among others.}
\label{datasphere_paper}
\end{figure}

In the few decades since its beginnings, the World Wide Web has embedded itself into the fabric of human society at a speed only matched by few technologies that came before it, and in the process, it has given an enormous boost to scientific endeavors across all disciplines. Particularly the discipline of web science---in which the web itself, as a socio-technical system, becomes the subject of inquiry---has proven a productive field of research~\cite{kilgarriff:2003,berners-lee:2006}. It is well understood that the vast size and sheer chaotic structure of the web pose a challenge to those that seek to study it~\cite{hendler:2008}; yet its precise dimensions are not at all easily quantified.

In what follows, we derive an informed estimate of the web's size and conclude that it represents, in fact, only a small fraction of all existing data. To this end, we start by considering the broader question of how much data there is overall. This question can be asked in two ways: First, how much data is generated continuously as a result of human activity; and secondly, how much data is persistently stored?
To answer the first question is to determine (according to the International Data Corporation) the size of the Global Datasphere, a ``measure of all new data captured, created, and replicated in a single year.''~\cite{idc:2018}.
Thanks to efforts at tracking the global market for computer storage devices, spearheaded by storage manufacturers and market researchers alike, we can make an educated guess at how much data that might be (see Figure~\ref{datasphere_paper} for an illustration).

Analyses like the ``Data Never Sleeps'' series~\cite{domo:2019,domo:2020} by business intelligence company Domo, Inc., provide pieces of the puzzle: While as recently as~2012, only a third of the world's population had internet access~\cite{domo:2017}, that figure has risen to 59\,\%, or 4.5~billion people, today.
Throughout~2018, the combined activity of internet users produced almost two petabytes per minute on average~\cite{domo:2019}. This year, in one minute of any given day, users upload 500~hours worth of video to YouTube, and more than a billion of them initiate voice or video calls~\cite{domo:2020}. This combined effort of internet users, however, is nowhere near the total volume of the Global Datasphere. In the same period, as automatic cameras record security footage and ATMs and stock markets transmit banking transactions, the ATLAS particle detector at CERN's Large Hadron Collider records a staggering amount of nearly four petabytes worth of particle collisions~\cite{brumfiel:2011}. 

Past and current projections by the aforementioned IDC complete our picture of the Global Datasphere's size. Already in 2014, more than four zettabytes of data were created yearly, and this amount has since increased at a rate of more than 150\,\% per year~\cite{idc:2014}. The total figure had grown to~33 zettabytes by~2018~\cite{idc:2018} and was previously projected to reach 44~zettabytes by~2020, and between~163 and 175~zettabytes by~2025~\cite{idc:2014,idc:2018,idc:2020a}. However, in the wake of the COVID-19 pandemic, the growth of the global datasphere has sharply accelerated and updated projections expect it to exceed~59~zettabytes (or $5.9\cdot 10^{22}$ bytes) already by the end of~2020~\cite{idc:2020c}.

To put this number into context, consider the following thought experiment: The largest commercially available spinning hard disk drives at the time of writing hold 18~terabytes in a~3.5-inch form factor. To store the entirety of the Global Datasphere would require more than 3.2~billion such disks, which---densely packed together---would more than fill the volume of the Empire State Building and almost cost the equivalent of Apple's two-trillion-dollar market capitalization in retail price.%
\footnote{We disregard issues like power, cooling, and redundancy, but also what should be a sizeable bulk discount.}
The densest SSD storage available today would reduce the volume requirements six-fold, while increasing the purchase cost by a factor of eight. However, even by the year~2025, four out of every five stored bytes are still expected to reside on rotating-platter hard drives~\cite{idc:2020a,idc:2020b}.

While all of the above makes the vastness of the Global Datasphere quite palpable, it leaves out one key point: Most of the 59~ZB is transient data that is never actually persistently stored. Once again, CERN's particle detectors quite impressively illustrate this: Almost all recorded particle collisions are discarded as uninteresting already in the detector, such that out of those four petabytes per minute, ATLAS outputs merely 19~gigabytes for further processing~\cite{brumfiel:2011}. Analogously, most data generated by and about internet users does not become part of the permanent web, which in and of itself is only a fraction of what resides on the world's storage media~\cite{idc:2014}.

\section{The Archived Web}

\begin{figure}[tb]
\centering
\includegraphics[width=\textwidth]{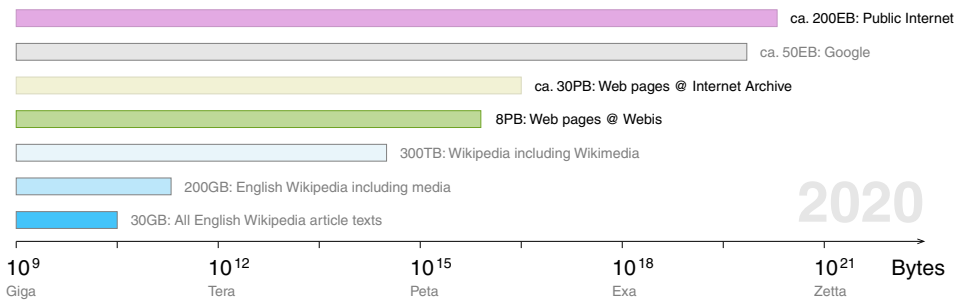}
\smallskip
\caption{The 2020 sizes of large, persistent data sets in comparison, illustrated on a logarithmic scale. The English Wikipedia is often compared to the Encyclop\ae{}dia Britannica, but largely exceeds it by a factor of more than~80 at a combined size of only 30\,GB. This is just over a one hundred millionth part of all textual web pages, whose total volume we estimate significantly smaller than 1\,EB (cf.\ \figurename~\ref{datasphere_paper}).}
\label{data-source-sizes_paper}
\end{figure}

Statista and Cisco Systems estimate that in the year~2020, about two zettabytes of installed storage capacity exist in datacenters across the globe~\cite{statista:2020a}. Worldwide storage media shipments support this claim. About 6.5\,ZB have been shipped cumulatively since 2010. Accounting for a yearly 3\,\% loss rate, as well as the fact that only a portion of all storage resides in enterprise facilities, leads to the same estimate of about 2--2.5\,ZB installed capacity in~2020~\cite{idc:2020a}. About half this capacity appears to be effectively utilized in datacenters, which means that the industry stores one zettabyte of data~\cite{statista:2020b}. Adding to this the remaining 30\,\% share of data from consumer devices, we can estimate that the world keeps watch over a data treasure of roughly~1.43 zettabytes in total at the time of writing. Although the majority of these sextillion stored bytes is probably quite recent~\cite{petrov:2020}, the figure does include all data stored (and retained) since the beginning of recorded history.
The majority of persistent data is not publicly accessible, but hides in closed repositories belonging to individuals and private or statutory organizations. But even of what we call the ``public internet,'' only the small ``surface web'' portion---openly accessible and indexed web pages, books, audiovisual media, and software---is actually {\it publicly} accessible. What remains is called the ``deep web,'' which includes email communications, instant messaging, restricted social media groups, video conferencing, online gaming, paid music and video-on-demand services, document cloud storage, productivity tools, and many more, but also the so-called ``dark web.'' The size of the deep web is much harder to assess than that of both the surface web and the world's global storage capacity. Folklore estimates range from 20~to 550~times the size of the surface web, but, to the best of our knowledge, no reliable sources exist and previous attempts at measuring its size typically relied on estimating the population of publicly retrievable text document IDs~\cite{lu:2010}. In absence of reliable data, we assume its actual size to be on the larger end and take an educated guess of up to 200~exabytes. That is a fifth of all persistent data and less than half a percent of the Global Datasphere~\cite{idc:2020a}.

The indexed surface web is substantially easier to quantify. In~2008, Google reported to have discovered one trillion distinct URLs~\cite{google:2008}, though only a portion of that was actually unique content worth indexing.
Based on estimates of the current index sizes of the two largest web search engines~\cite{van-den-bosch:2016}, there are approximately~60 billion pages (lower bound: 5.6~billion) in the indexed portion of the World Wide Web as of~2020.%
\footnote{\url{https://www.worldwidewebsize.com}}
The HTTP Archive,%
\footnote{\url{https://httparchive.org}}
which aims to track the evolution of the web from a metadata perspective, records (among other metrics) statistics about page weights, i.e., the total number of bytes transferred during page load. Based on the page weight percentiles given in their~2019 report~\cite{everts:2019}, we estimate that the combined size of all indexed web pages is at least~1.8 petabytes counting their HTML payload alone. As a rather generous upper bound, we assert that the World Wide Web including all directly embedded style, script, and image assets is (not accounting for redundancy) most likely less than 200~petabytes and---considering the English-language bias~\cite{van-den-bosch:2016} of the employed method---certainly less than one exabyte in size~(Figure~\ref{datasphere_paper}). As such, the surface web accounts for less than half a percent of the presumed 200\,EB constituting the public internet.
It is important to note that this figure only captures the {\it textual} part of the publicly indexable surface web. Publicly accessible on-demand video streaming, accounting for 15\,\% of all downstream internet traffic in 2020~\cite{sandvine:2020} (60\,\% including paid subscriptions) and necessarily taking a sizable portion of enterprises' mass storage, is not included.

Figure~\ref{data-source-sizes_paper} compares the sizes of large, persistent data sets on the public internet, starting with an estimate for the total amount of data stored by Google. While Google does not publish any official figures on their data centers' storage capacity, they have maintained a public list of datacenter locations for at least the past eight years~\cite{google:2013}. In~2013, thirteen datacenters were listed worldwide and contemporary estimates put the total amount of data stored by Google at 15\,EB~\cite{munroe:2013}. By mid-2020, Google's list of datacenters had grown to~21 entries~\cite{google:2020}. At the same time, the areal density of hard disk storage has approximately doubled since~2013~\cite{coughlin:2020}. For the total amount of data stored by Google in~2020, we hence propose an updated estimate of~50\,EB---this includes both public and non-public data.

The Internet Archive%
\footnote{\url{https://archive.org}}
is a large-scale, nonprofit digital library covering a wide range of media, including websites, books, audio, video, and software. In~2009, the Internet Archive's entire inventory comprised only one petabyte~\cite{jaffe:2009}, but according to current statistics,%
\footnote{\url{https://archive.org/~tracey/mrtg/du.html}} %
it has grown by a factor of almost~100 over the past decade. The Wayback Machine, an archive of the World Wide Web, forms a significant portion of the Internet Archive's collections. As of this writing, it comprises nearly~500 billion web page captures, which take up approximately~30 petabytes of storage space. In the Webis research group, we aim to store up to~8\,PB of web archive data on our own premises, much of it originating from the Internet Archive, but also from other sources, such as the Common Crawl. 

The Wikipedia is commonly used in large-scale text analytics tasks and thus we include it in the comparison in three ways: The combined size of all Wikipedias, including the files in the Wikimedia Commons project, comes at approximately~300\,TB,%
\footnote{\url{https://commons.wikimedia.org/wiki/Special:MediaStatistics}}
of which approximately~3\,TB are actual wikitext.%
\footnote{\url{https://stats.wikimedia.org/\#/all-projects/content/net-bytes-difference/normal|bar|all|~total|monthly}}
The English Wikipedia including all media files is three orders of magnitude smaller, at about~200\,GB.%
\footnote{\url{https://en.wikipedia.org/wiki/Special:MediaStatistics}}
Lastly, the plain wikitext content of the English Wikipedia is yet another order of magnitude smaller, adding up to merely~30\,GB in~2020.%
\footnote{\url{https://en.wikipedia.org/w/index.php?title=Wikipedia:Size_in_volumes&oldid=966973828}}

\begin{figure}[t]
\centering
\includegraphics[width=\textwidth]{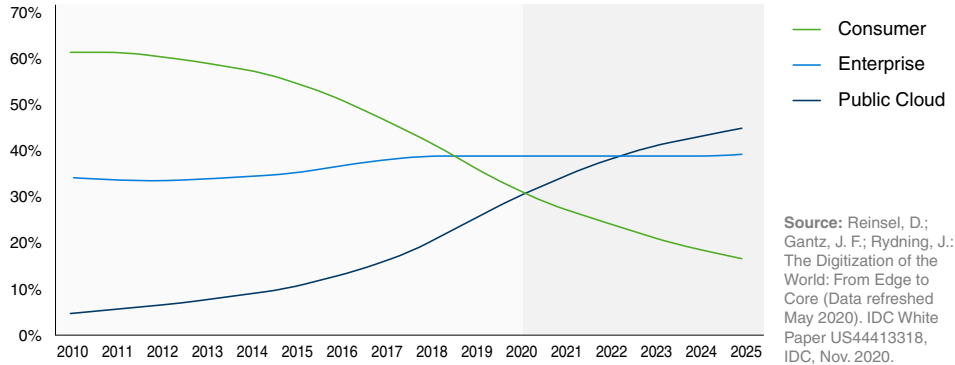}
\smallskip
\caption{As of 2020, IDC believes, more bytes are stored in the public cloud than on consumer devices. By 2022, more data is going to be stored in the cloud than in traditional datacenters.}
\label{idc18-where-is-the-data-stored_paper}
\end{figure}

Figure~\ref{idc18-where-is-the-data-stored_paper} is based on an analysis by IDC and Seagate~\cite{idc:2018,idc:2020a} inquiring \emph{where} the world's persistent data has been, and is going to be stored in the future: A decade ago, most of the data created across the world remained on largely disconnected consumer endpoint devices, but that has steadily shifted towards public clouds. As of 2020, the share of data stored in public clouds has exceeded that stored on consumer devices, and within the next three years, the majority of persistent data is predicted to reside in public clouds.

The rapid growth in both size and publicness of web-based data sources drives interest in web science globally, not only in the Webis group. We have shown the persistent World Wide Web to be quite a bit smaller than one might expect in contrast with the totality of data created and stored; nevertheless, true ``web archive analytics'' tasks, which we still consider to involve datasets exceeding hundreds of terabytes, certainly do require a large and specialized hardware infrastructure to be feasible.

\section{Web Archive Analytics at Webis}

\begin{figure}[t]
\centering
\includegraphics[width=\textwidth]{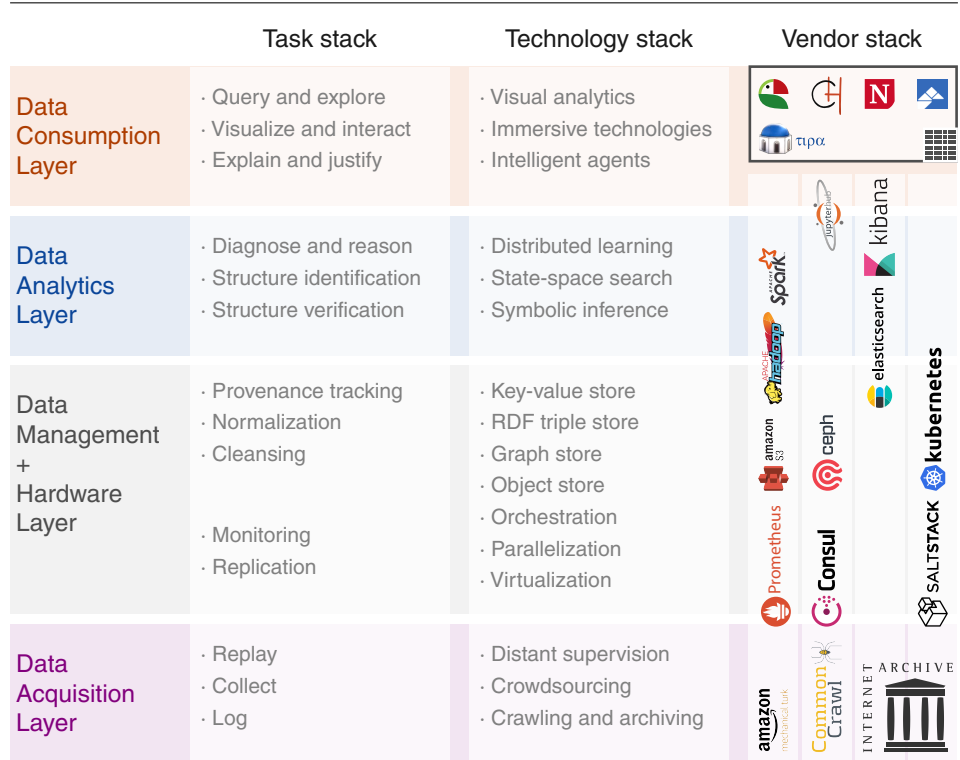}
\smallskip
\caption{The Webis web archive analytics infrastructure stack. The $y$-axis organizes (bottom to top) the processing pipeline from data acquisition to data consumption. The $x$-axis organizes (right to left) the employed tools (vendor stack), algorithms and methods (technology stack), and tackled problem classes (task stack). The box at the top of the vendor stack shows the resulting frontend services maintained by the Webis research group: \args, \chatnoir, \netspeak, \picapica, and \tira.}
\label{webis-analytics-stack_paper}
\end{figure}

Figures~\ref{hardware-webis-2009-2020bd_paper} and~\ref{webis-analytics-stack_paper} illustrate the software and infrastructure relevant to this effort: Two different computing clusters contribute to our web archive analytics efforts, out of five total currently operated by our research group. We have maintained cluster hardware in some form or another for more than a decade. Figure~\ref{hardware-webis-2009-2020bd_paper} outlines their specifications. In the following, we give a short overview of typical workloads and research areas each individual cluster is tasked with: \alphaweb is made up of repurposed desktop hardware and enabled our earliest inroads into web mining and analytics; nowadays it still functions as a teaching and staging environment. \betaweb is currently our primary environment for web mining, batch and stream processing, data indexing, virtualization, and the provision of public web services. \gammaweb's primary purpose is the training of machine learning models, text synthesis, and language modeling. \deltaweb is focused on storage and serves as a scalable and redundant persistence backend for web archiving and other tasks. \epsilonweb is used for indexing and argument search.

The \betaweb and \deltaweb clusters are the primary workhorses in our ongoing web archiving and analytics efforts. The clusters comprise~135 and 78~Dell PowerEdge servers, respectively, spread across two datacenters joined via a 400\,GB/s interconnect. Each individual node is attached to one of two middle-of-row Cisco Nexus switches via a 10\,GB/s link. The storage cluster maintains more than 12\,PB of raw storage capacity across 1\,248 physical spinning disks. The compute cluster possesses an additional 4.3\,PB for serving large indexes and storing ephemeral intermediate results across 1\,080 physical disks.

\begin{figure}[t]
\centering
\includegraphics[width=\textwidth]{hardware-webis-2009-2020bd_paper}
\smallskip
\caption{The cluster hardware owned by the Webis research group, organized from left to right by acquisition date (shown in square brackets). \betaweb and \deltaweb have been specifically designed to handle large-scale web archive analytics tasks.}
\label{hardware-webis-2009-2020bd_paper}
\end{figure}

Our analytics stack is shown conceptually in Figure~\ref{webis-analytics-stack_paper}, beginning with the bottom-most data acquisition layer: Primary sources for data ingestion include web crawls and web archives, such as the aforementioned Internet Archive, the Common Crawl,%
\footnote{\url{https://commoncrawl.org}}
the older ClueWeb%
\footnote{\url{https://lemurproject.org}}
datasets, as well as our own specialized crawls focused on individual research topics, such as argumentation and authorship analytics. Crowdsourcing services, such as Amazon Mechanical Turk, support and complement (distantly) supervised learning tasks.

Both the \betaweb and \deltaweb cluster are provisioned and orchestrated using the SaltStack configuration management and IT automation software, which manages the deployment and configuration of fundamental file-system- and infrastructure-level services on top of a network-booted Ubuntu Linux base image. We supervise the proper functioning of all system components with the help of a Consul cluster (for service discovery) and a redundant Prometheus deployment (for event monitoring and alerting).

The primary purpose of the \deltaweb cluster is to serve a distributed Ceph~\cite{weil:2007} storage system with one object storage daemon (OSD) per physical disk, as well as five redundant Monitor (MON) / Manager (MGR) daemons. The bulk of the data payload is accessed via the RadosGW S3 API gateway service, which is provided by seven redundant RadosGW daemons and backed by an erasure-coded storage pool. Smaller and more ephemeral datasets live in a CephFS distributed file system with three active and four standby Metadata Servers~(MDS).

On \betaweb, we maintain a Saltstack-provisioned Kubernetes cluster on top of which most of our internal and public-facing services are deployed. Kubernetes services are provided with persistent storage through a RADOS Block Device (RBD) pool in the Ceph cluster. A Hadoop Distributed File System (HDFS) holds temporary intermediate data.

Services operated on top of Kubernetes include a 130-node Elasticsearch deployment powering our search engines, as well as a suite of data analytics tools such as Hadoop Yarn, Spark, and JupyterHub. Our public-facing services are, among others, the search engines \args~\cite{stein:2017r,stein:2019k} and \chatnoir~\cite{stein:2018c}, the \netspeak~\cite{stein:2010b,stein:2014d} writing assistant, the \picapica~\cite{stein:2017v} text reuse detection system, and the \tira~\cite{stein:2012k,stein:2019r} evaluation-as-a-service platform. The web-archive-related services at \url{archive.webis.de} are still largely under construction. Ultimately, we envision also allowing external collaborators to process and analyze web archives on our infrastructure. As of October~2020, almost~2.3\,PB of data---of which 560\,TB stem from the Internet Archive and the rest from the Common Crawl---have been downloaded and are stored on our infrastructure.

\section{Conclusion}

The web and its ever-growing archives represent an unprecedented opportunity to study humanity's cultural output at scale. While in 2020 alone, a staggering~59 zettabytes of data will have been created throughout the year, most of that is transient, and the World Wide Web is nearly five orders of magnitude smaller. Nevertheless, the scale of the web is large enough to be challenging, and so web analytics requires potent and robust infrastructure.

The Webis cluster infrastructure has already enabled many research projects on a scale which typically only big industry players can afford due to the massive hardware requirements. This allows us to develop and maintain services such as the \chatnoir web search engine~\cite{stein:2018c}, the argument search engine \args~\cite{stein:2017r}, but also facilitates individual research projects which rely on utilizing the bundled computational power and storage capacity of the capable Hadoop, Kubernetes, and Ceph clusters. A few recent examples are the analysis of massive query logs~\cite{stein:2020a}, a detailed inquiry into near-duplicates in web search~\cite{webis:2020b,webis:2020d}, the preparation, parsing, and assembly of large NLP~corpora~\cite{stein:2020o,stein:2020u}, and the exploration of large transformer models for argument retrieval~\cite{akiki:2020}.

With significant portions of the Internet Archive available and further hardware acquisitions in progress, many more of these research projects at a very large and truly representative scale will become feasible, including not only the evaluation, but also the efficient training of ever larger machine learning models.


\end{document}